\documentclass{article}

\usepackage{PRIMEarxiv}

\usepackage[utf8]{inputenc} % allow utf-8 input
\usepackage[T1]{fontenc}    % use 8-bit T1 fonts
\usepackage{hyperref}       % hyperlinks
\usepackage{url}            % simple URL typesetting
\usepackage{booktabs}       % professional-quality tables
\usepackage{amsfonts}       % blackboard math symbols
\usepackage{nicefrac}       % compact symbols for 1/2, etc.
\usepackage{microtype}      % microtypography
\usepackage{lipsum}
\usepackage{fancyhdr}       % header
\usepackage{graphicx}       % graphics
\graphicspath{{media/}}     % organize your images and other figures under media/ folder
\usepackage{amsmath}
\usepackage{caption}
\usepackage{subcaption}
\title{TPDH-Graphene as a New Anodic Material for Lithium Ion Battery: DFT-Based Investigations
} 

\author{
  Juan Gomez Quispe, Bruno Bueno Ipaves Nascimento, Pedro Alves da Silva Autreto \\
  Center of Natural and Human Sciences, Federal University of ABC \\
  Santo Andre \\
  Sao Paulo, Brazil\\
  \texttt{pedro.autreto@ufabc.edu.br} \\
   \And
   \\ Douglas Soares Galvao \\ 
   \texttt{galvao@ifi.unicamp.br} \\
  Applied Physics Department, State University of Campinas, Campinas, Sao Paulo, Brazil \\
}

\begin{document}
\maketitle

%%%%%%%%%%%%%%%%%%%%%%%%%%%%%%%%%%%%%%%%%%%%%%%%%%%%%%%%%%%%%%%%%%%%%
%% The "tocentry" environment can be used to create an entry for the
%% graphical table of contents. It is given here as some journals
%% require that it is printed as part of the abstract page. It will
%% be automatically moved as appropriate.
%%%%%%%%%%%%%%%%%%%%%%%%%%%%%%%%%%%%%%%%%%%%%%%%%%%%%%%%%%%%%%%%%%%%%
%\begin{tocentry}

%Some journals require a graphical entry for the Table of Contents.
%This should be laid out ``print ready'' so that the sizing of the
%text is correct.

%Inside the \texttt{tocentry} environment, the font used is Helvetica
%8\,pt, as required by \emph{Journal of the American Chemical
%Society}.

%The surrounding frame is 9\,cm by 3.5\,cm, which is the maximum
%permitted for  \emph{Journal of the American Chemical Society}
%graphical table of content entries. The box will not resize if the
%content is too big: instead it will overflow the edge of the box.

%This box and the associated title will always be printed on a
%separate page at the end of the document.

%\end{tocentry}

%%%%%%%%%%%%%%%%%%%%%%%%%%%%%%%%%%%%%%%%%%%%%%%%%%%%%%%%%%%%%%%%%%%%%
%% The abstract environment will automatically gobble the contents
%% if an abstract is not used by the target journal.
%%%%%%%%%%%%%%%%%%%%%%%%%%%%%%%%%%%%%%%%%%%%%%%%%%%%%%%%%%%%%%%%%%%%%
\begin{abstract}
  The potential of tetra-penta-deca-hexagonal graphene (TPDH-gr), a recently proposed 2D carbon allotrope as an anodic material in lithium ion batteries (LIB), was investigated through density functional theory (DFT) calculations. The results indicate that Li-atom adsorption is moderate (around 0.70 eV), allowing for easy desorption. Moreover, energy barrier, diffusion coefficient, and open circuit voltage (OCV) calculations show rapid Li atom diffusion on the TPDH-gr surface, stable intercalation of lithium atoms, and good performance during the charge and discharge cycles of the LIB. These findings, combined with the intrinsic metallic nature of TPDH-gr, indicate that this new 2D carbon allotrope is a promising candidate for use as an anodic LIB material.
  
\end{abstract}

%%%%%%%%%%%%%%%%%%%%%%%%%%%%%%%%%%%%%%%%%%%%%%%%%%%%%%%%%%%%%%%%%%%%%
%% Start the main part of the manuscript here.
%%%%%%%%%%%%%%%%%%%%%%%%%%%%%%%%%%%%%%%%%%%%%%%%%%%%%%%%%%%%%%%%%%%%%
\section{Introduction}

Graphene, a revolutionary material, is a two-dimensional (2D) structure composed of a single layer of carbon atoms arranged in a hexagonal honeycomb lattice, where carbon atoms are covalently bonded in $sp^{2}$ hybridization \cite{Geim2009}. This hybridization results in excellent mechanical rigidity and high electronic mobility due to the sigma and $\pi$ type bonds \cite{Zhang2022c}. 

In contrast to graphene, other 2D carbon allotropes can be formed by combining different hybridization states ($sp$, $sp^{2}$ or $sp^{3}$) and different types of carbon rings. One example is graphenylene, which is composed of tetragonal ($C_{4}$), hexagonal ($C_{6}$), and dodecagonal ($C_{12}$) carbon rings \cite{Jana2022}. Calculations based on density functional theory (DFT) indicate that graphenylene has a semiconducting nature with a direct band gap value of approximately 25 meV \cite{Song2013}. Another carbon allotrope, recently synthesized \cite{Fan2019c}, is TPH-graphene, which is formed by $C_{4}$, $C_{5}$, and $C_{6}$ carbon rings, which can exist in two different semiconducting phases with direct band gap values of 2.704 and 2.361 eV, respectively \cite{Zhang2021c}. Much of the research related to these new 2D carbon allotropes aims at studying their mechanical and electronic properties and their potential application as an anode material in lithium-ion batteries (LIB).\\

Y. Yu investigated the application of graphenylene as an anodic material in LIBs using DFT calculations \cite{Yu2013}. The adsorption of Li atoms in graphenylene was reported to be stronger than in pristine graphene, achieving a theoretical capacity of 1116 $\text{mA h g}$$^{-1}$, which is greater than the graphite capacity. Furthermore, due to the carbon rings $C_{12}$, diffusion paths were found with low-energy barriers ($< 0.9$ eV), making it possible to achieve that lithium atoms can be well dispersed on the graphenylene. Recently, T. Qiu et al. \cite{Qiu2022}, reported a theoretical capacity of 837 $\text{mA h g}$$^{-1}$ and low diffusion barriers ($< 0.83$ eV) for a new 2D carbon allotrope composed of Quadrangular, Pentagonal, Hexagonal rings and large Tetradecagonal pores (QPHT-graphene) and that exhibit metallic behavior with a significant number of flat bands near the Fermi level region \cite{Qiu2022,Jana2022}.\\

\begin{figure}[t]
\centering
\begin{subfigure}{0.32\textwidth}
    \includegraphics[width=\textwidth]{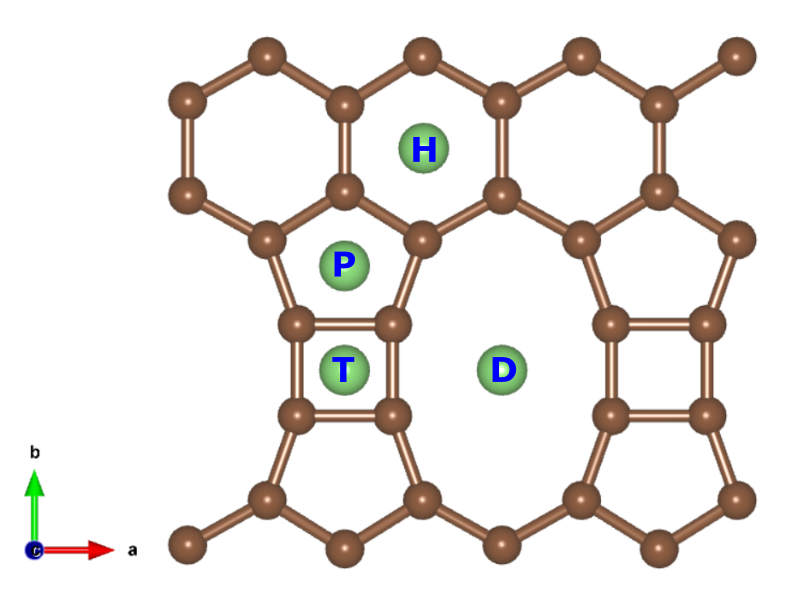}
    \caption{}
    \label{fig:1a}
\end{subfigure}
\hfill
\begin{subfigure}{0.32\textwidth}
    \includegraphics[width=\textwidth]{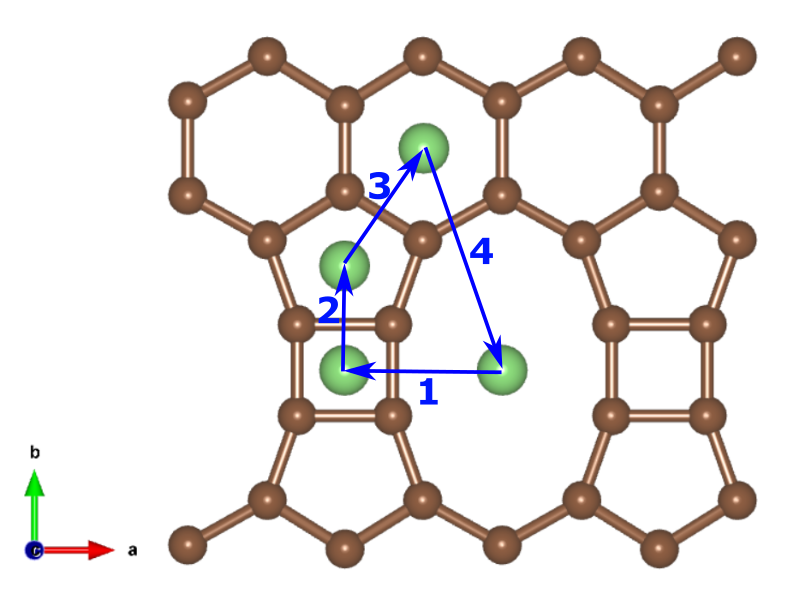}
    \caption{}
    \label{fig:1b}
\end{subfigure}
\hfill
\begin{subfigure}{0.32\textwidth}
    \includegraphics[width=\textwidth]{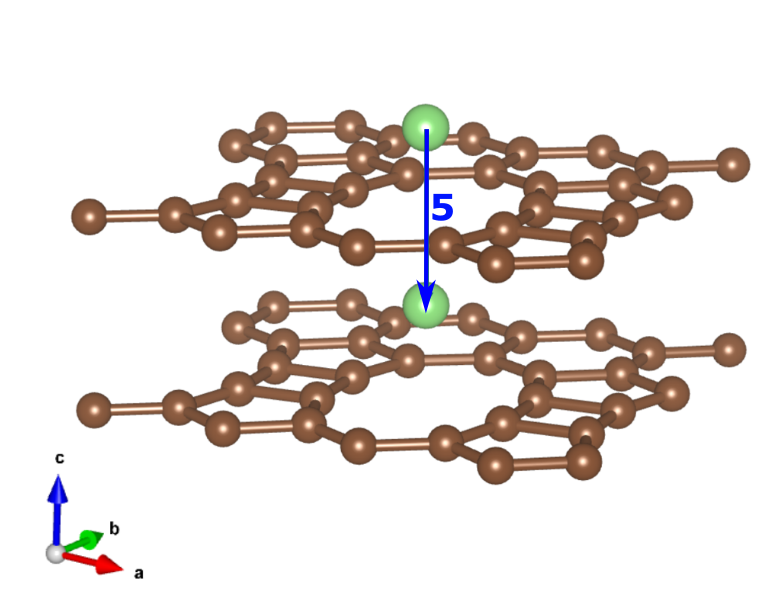}
    \caption{}
    \label{fig:1c}
\end{subfigure}
\caption{(a) Selected TPDH-gr regions for Li adsorption. (b) Investigated diffusion paths of Lithium ions on the TPDH-gr monolayer. (c) Investigated diffusion path in a TPDH-gr bilayer. These paths were chosen based on minimum energy pathways. See text for discussions.}
\label{fig:figures}
\end{figure}

The essential requirements for the application of a material as an anode of a LIB are: good thermal stability to not induce structural deformations, high porosity values that are generally satisfied by high-order carbon rings ($C_{> 10}$), and high values for electrical conductivity \cite{Ma2021}. The vast majority of these new 2D carbon allotropes have good thermal stability, because of strong C-C covalent bonds and high porosity values. However, they can also have large electronic band gap values, negatively influencing their application in LIBs. For example, pentagraphene, formed by $C_{5}$ rings, has a theoretical capacity of 1489 $\text{mA h g}$$^{-1}$, but has a band gap value of $\sim$ 3.25 eV \cite{Xiao2016}. Furthermore, a theoretical capacity of 3916 $\text{mA h g}$$^{-1}$ was reported for twin-graphene, formed by $C_{6}$ rings, this value being the largest reported so far \cite{Rajkamal2019}. Twin-graphene has semiconductor behavior with a band-gap value of 0.75 eV \cite{Gao2023}.\\

Tetra-penta-deca-hexagonal-graphene (TPDH-gr) was recently proposed by D. Bhattacharya and D. Jana \cite{Bhattacharya2021}. It is a 2D carbon allotrope composed of fully $sp^{2}$ carbon atoms arranged in $C_{4}, C_{5}, C_{10}$ and $C_{6}$ rings. TPDH-gr exhibits good mechanical and thermal stability up to 1000K. In addition, TPDH-gr exhibits anisotropic elastic properties due to its ring arrangement topology. The $C_{4}$ rings of TPDH-gr were recently reported to be more reactive, capable of adsorbing a larger amount of hydrogen compared to their other rings \cite{D3CP00186E}. Furthermore, Oliveira et al. demonstrated that hydrogenation of TPDH-gr in the $C_{4}$ rings also results in anisotropic thermoelectric properties \cite{C.Oliveira2024SelectiveTPDH-Graphene}.\\

In this work, the potential application of TPDH-gr as a LIB anode material was investigated using density functional theory (DFT)-based simulations. Through adsorption calculations, a high theoretical capacity of 1116 $\text{mA h g}$$^{-1}$ for TPDH-gr was estimated, and lower diffusion barriers ($< 0.2$ eV) were found for Li atoms, comparable to the energy barriers of graphite, which is the most commercially used LIB anode.

\section{Computational Methods}

The SIESTA software \cite{Soler2002a,Garcia2020} was utilized for all DFT simulations. For all calculations, a $2 \times 1$ supercell of TPDH-gr was considered. Periodic boundary conditions were applied to mimic infinitely large systems. A vacuum space of $20$ \AA \ was set to prevent spurious interactions between the layer and its periodic images. Within SIESTA, the Kohn-Sham orbitals were expanded using a double-$\zeta$ basis set consisting of numerical pseudoatomic orbitals of finite range, augmented with polarization orbitals. Optimized pseudopotentials and atomic bases from the SIMUNE database \cite{SIMUNE} were employed with the Perdew-Burker-Ernzerhof (PBE) approximation for the exchange and correlation functional. Furthermore, van der Waals interaction corrections, equivalent to DF1 \cite{Dion2004}, were incorporated into standard DFT calculations. The Brillouin Zone (BZ) sampling utilized a 5 $\times$ 8 $\times$ 1 irreducible Monkhorst-Pack (MP) k-point grid \cite{Hu2019}. The total energy convergence threshold for each electronic calculation was set at $1 \times 10^{-4}$ eV. Geometry optimizations were performed using the conjugate gradient (CG) algorithm, ensuring that the magnitude of total forces acting on each ion was minimized to less than 0.01 \AA/eV via ionic position displacements. \\

We calculated the adsorption energy ($E_{\text{ads}}$) of a lithium atom on TPDH-gr layer using the following equation:
\begin{equation}
 E_{\text{ads}} = E_{\text{Li + TPDH-gr}} - (E_{\text{Li}} + E_{\text{TPDH-gr}}),
\end{equation}
where $E_{\text{Li + TPDH-gr}}$ is the total energy of the final configuration of the Li atom adsorption process, while $E_{\text{Li}}$ and $E_{\text{TPDH-gr}}$ are the total energy of a gas system of Li and TPDH-gr. An alternative definition of the adsorption energy $E_{ads}^{bcc}$ was also considered, where the total energy per atom of lithium in bulk form (bcc) is considered. For both definitions of adsorption energy, negative values indicate that the Li atom is energy favorably adsorbed on TPDH-gr, while positive values indicate no adsorption. As shown in Figure \ref{fig:1a}, four adsorption regions (centers of the different rings) were considered: tetragonal (T), pentagonal (P), decagonal (D), and hexagonal (H) sites.\\

Adsorption calculations were carried out for the (N = 1,2,...,6) Li atoms to find the storage capacity limit of TPDH-gr. For each value of N, $10$ random different configurations were considered in which the distances between the lithium atoms were greater than $2.80$ \AA \, which is slightly larger than the molecular distance of $\text{Li}_{2}$ (2.60 \AA ), to prevent Li clustering \cite{Ullah2024}. These new $E_{ads}(N)$ are calculated using the following formula:
\begin{equation}
 E_{\text{ads}}(N) = E_{\text{N-Li + TPDH-gr}} - (N E_{\text{Li}} + E_{\text{TPDH-gr}}),
\end{equation}
where $E_{\text{N-Li + TPDH-gr}}$ is the total energy of TPDH-gr with $N$ adsorbed Li atoms. Through these calculations, it is possible to determine the maximum number of $N$ Li atoms that can be absorbed, which is where the adsorption energy becomes positive \cite{Ullah2024}. Therefore, once the value of $N$ has been determined, it is easy to estimate the theoretical capacity of TPDH-gr using the following formula:
\begin{equation}
    C = N_{\text{max}} \times \frac{F}{M},
    \label{eq:capacity}
\end{equation}
where $N_{\text{max}}$ is the maximum number of Li atoms adsorbed on TPDH-gr, $F = 96485.332$ s A/mol is the Faraday constant, and $M$ is the molar mass of the TPDH-gr supercell.\\

\begin{figure}[b]
\centering
\begin{subfigure}{0.49\textwidth}
    \includegraphics[width=\textwidth]{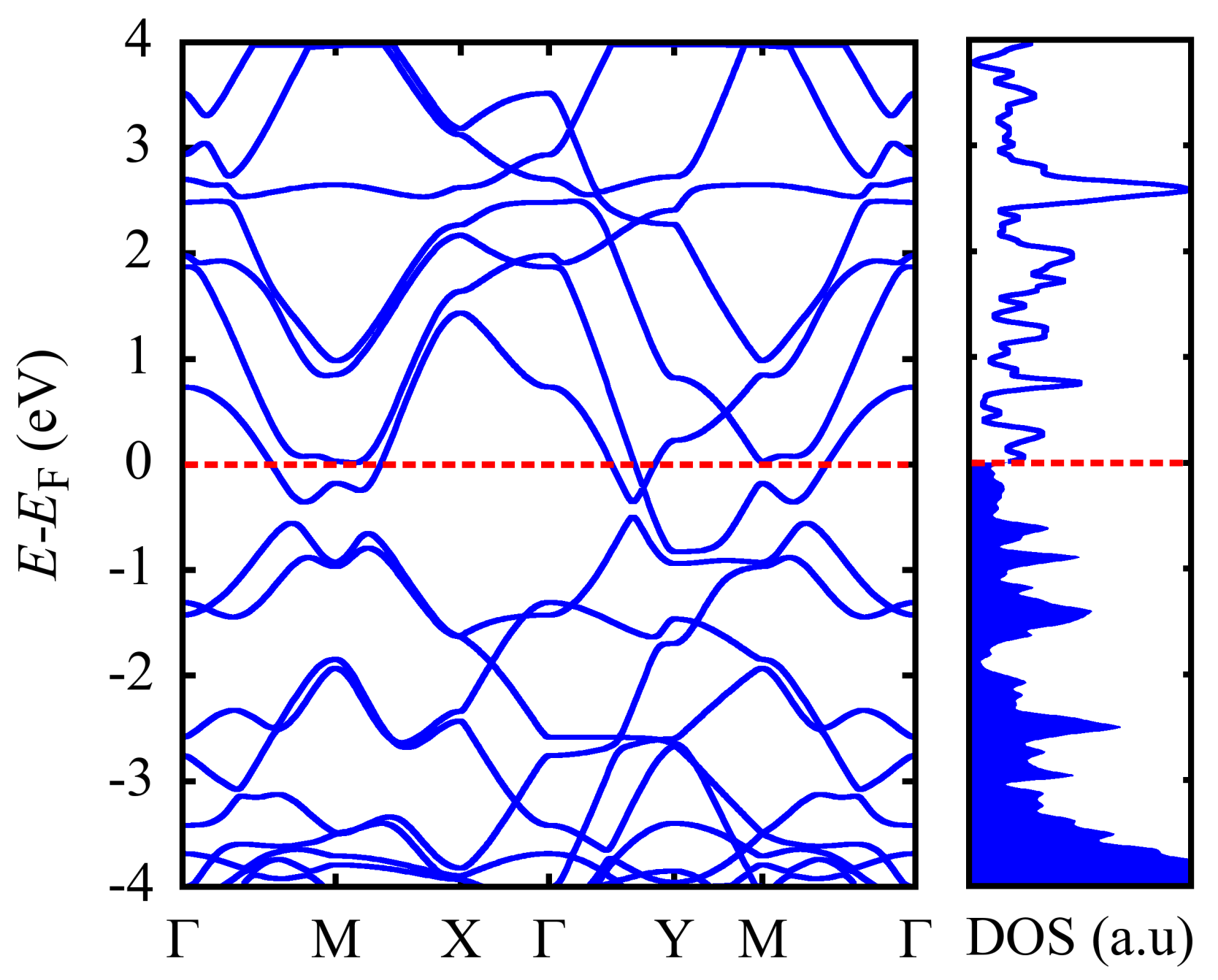}
    \caption{}
    \label{fig:2a}
\end{subfigure}
\hfill
\begin{subfigure}{0.49\textwidth}
    \includegraphics[width=\textwidth]{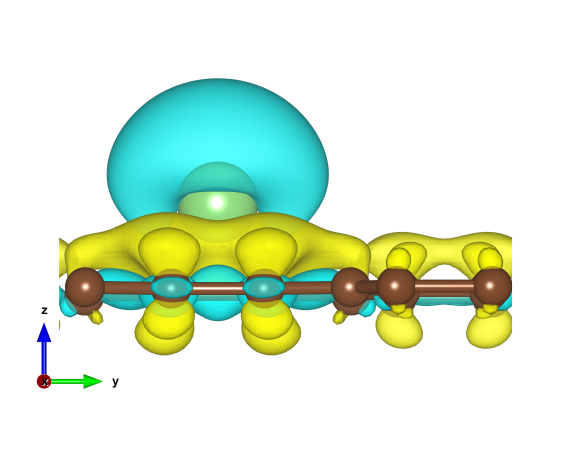}
    \caption{}
    \label{fig:2b}
\end{subfigure}
\caption{(a) TPDH-gr electronic band structure after the adsorption of Li on the D region. (b) Charge density difference for the adsorption of a single Li atom on TPDH-gr in the D region. The cyan and yellow regions represent the electron losses and gains, respectively. Here, the net charge density difference is defined as the Li-adsorbed TPDH-gr charge density minus the isolated Li atom and TPDH-gr charge density.}
\label{fig:figures}
\end{figure}

Another important feature of an anode material is the average open-circuit voltage (OCV), which is generally defined as the voltage between the terminals of an electrochemical cell when no current flows through the cell. The OCV can be calculated using the following equation:\\

\begin{equation}
 \text{OCV} = \frac{(E_{\text{TPDH-gr}} + N E_{\text{Li}} - E_{\text{N-Li + TPDH-gr}})}{N e}
 \label{eq:ocv}
\end{equation}
where $e$ is the electronic charge of the Lithium-ion. As seen in the previous equation, the OCV is related to the adsorption energy of the lithium atom. Thus, OCV negative values indicate that the adsorption is unfavorable, and the Li atoms will tend to form clusters. In contrast, positive values indicate the possibility of intercalation of Lithium atoms, which further suggests good cycling performance \cite{He2019a,Kingori2021,Ipaves2022}.\\
 
To characterize Li diffusion on the surface of the TPDH-gr sheet, the minimum energy pathway (MEP) was calculated using nudged elastic band (NEB) calculations \cite{Sheppard2008,Smidstrup2014}. The NEB method was discretized into five images in the configuration space, with fictitious springs connected between the images to prevent them from converging to the same local minima. Initially, images were linearly interpolated between the reactive and product states of the reaction and then optimized using a conjugate gradient method. A climbing image scheme (CI-NEB) \cite{Henkelman2000} was implemented to ensure an accurate determination of the transition state (TS). Five diffusion paths for the Li atom were considered, as shown in Figures \ref{fig:1b} and \ref{fig:1c}.

\section{Results and discussion}

In Table \ref{table:adsoprtion_values}, we present the values of $E_{ads}$, $E_{ads}^{bcc}$, and the perpendicular distance $d_{ads}$ from the Li adsorbed atom to the TPDH-gr layer. The adsorption calculations were carried out for a monolayer and a bilayer of TPDH-gr. These results show that the adsorption of Li on TPDH-gr is thermodynamically favorable, with negative values, and that the sites D and P are the preferential adsorption regions. On the other hand, the values of $E_{ads}^{bcc}$ are also negative values; however, these values are smaller compared to the values of $E_{ads}$. Additionally, the TPDH-gr bilayer presents adsorption energy slightly higher than that of the monolayer. However, it is also shown that the vdW correction (DF1) has a small effect on the Li adsorption energies and geometries.\\

\begin{figure}[t]
\centering
\includegraphics[scale=0.8]{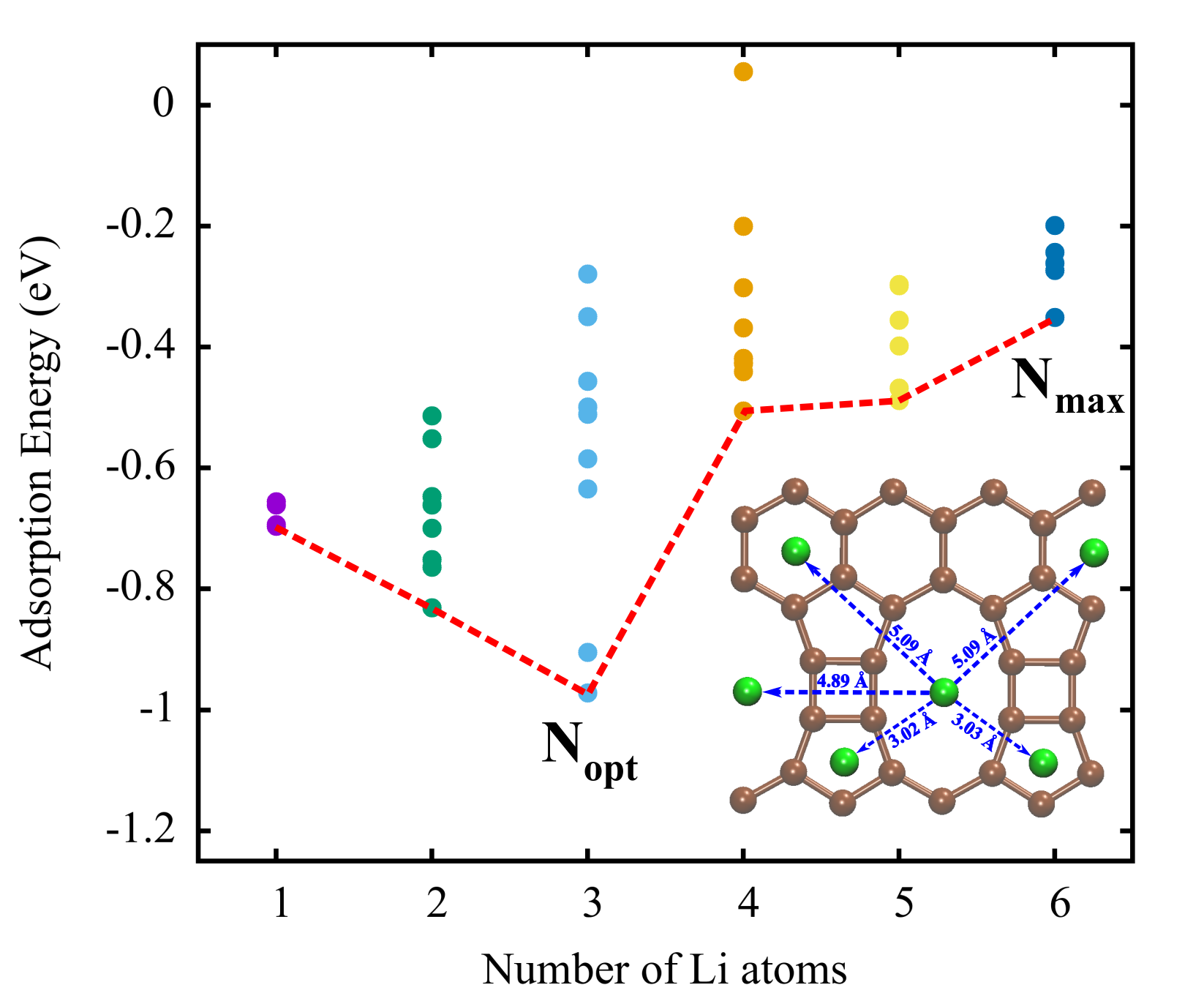}
\caption{Adsorption energies $E_{ads}^{bcc}$ (eV) as a function of the number N (N=1,2,...,6) of Li atoms. Inset figure: Configuration of TPDH-gr with N=6 Li atoms, showing that the distance between Li atoms is $> 2.80$ Å to prevent clustering.}
\label{fig:3a}
\end{figure}

\begin{table}[t]
\centering
\caption{\small{Adsorption energy values ($E_{ads}$ and $E_{ads}^{bcc}$) and final distance ($d_{ads}$) of the Li atom at the different adsorption regions (D, T, P, and H). For the adsorption process, TPDH-gr monolayers and bilayers were considered. van der Waals corrections (DF1) were also considered. The energies are given in eV, and the distances are in {\AA} }}
\begin{tabular}{llllllllll}
\hline
              & \multicolumn{4}{c}{\textbf{monolayer}}                                                               &                 & \multicolumn{4}{c}{\textbf{bilayer}}                                                                                  \\ \hline
              & \multicolumn{1}{c}{D}     & \multicolumn{1}{c}{T}     & \multicolumn{1}{c}{P}     & \multicolumn{1}{c}{H} & & \multicolumn{1}{c}{D}     & \multicolumn{1}{c}{T}     & \multicolumn{1}{c}{P}     & \multicolumn{1}{c}{H} \\ 
$E_{ads}$       & \multicolumn{1}{l}{-2.59} & \multicolumn{1}{l}{-2.55} & \multicolumn{1}{l}{-2.59} & -2.55             &     & \multicolumn{1}{l}{-2.65} & \multicolumn{1}{l}{-2.63} & \multicolumn{1}{l}{-2.67} & -2.63                  \\ 
$E_{ads}^{bcc}$  & \multicolumn{1}{l}{-0.70} & \multicolumn{1}{l}{-0.66} & \multicolumn{1}{l}{-0.69} & -0.66            &      & \multicolumn{1}{l}{-0.76} & \multicolumn{1}{l}{-0.74} & \multicolumn{1}{l}{-0.78} & -0.74                  \\ 
$d_{ads}$         & \multicolumn{1}{l}{1.46}  & \multicolumn{1}{l}{1.90}  & \multicolumn{1}{l}{1.82}  & 1.77         &          & \multicolumn{1}{l}{1.46}  & \multicolumn{1}{l}{1.86}  & \multicolumn{1}{l}{1.82}  & 1.76                   \\ 
%              & \multicolumn{4}{c}{\textbf{monolayer}}                                                                                & \multicolumn{4}{c}{\textbf{bilayer}}                                                                                  \\ 
%\textbf{vdw}              & \multicolumn{1}{c}{D}     & \multicolumn{1}{c}{T}     & \multicolumn{1}{c}{P}     & \multicolumn{1}{c}{H} & & \multicolumn{1}{c}{D}     & \multicolumn{1}{c}{T}     & \multicolumn{1}{c}{P}     & \multicolumn{1}{c}{H} \\ 
   %   &  &  &  &        &          &  &  &  &                 \\ 

$E_{ads}$ (DF1)       & \multicolumn{1}{l}{-2.35} & \multicolumn{1}{l}{-2.30} & \multicolumn{1}{l}{-2.32} & -2.27        &          & \multicolumn{1}{l}{-2.44} & \multicolumn{1}{l}{-2.41} & \multicolumn{1}{l}{-2.43} & -2.39                  \\ 
$E_{ads}^{bcc}$ (DF1)  & \multicolumn{1}{l}{-0.68} & \multicolumn{1}{l}{-0.63} & \multicolumn{1}{l}{-0.65} & -0.60       &           & \multicolumn{1}{l}{-0.77} & \multicolumn{1}{l}{-0.74} & \multicolumn{1}{l}{-0.76} & -0.72                  \\ 
$d_{ads}$ (DF1) & \multicolumn{1}{l}{1.38}  & \multicolumn{1}{l}{1.91}  & \multicolumn{1}{l}{1.86}  & 1.81           &        & \multicolumn{1}{l}{1.49}  & \multicolumn{1}{l}{1.95}  & \multicolumn{1}{l}{1.88}  & 1.83                   \\ \hline 

\end{tabular}
\label{table:adsoprtion_values}
\end{table}

\begin{figure}[t]
    \centering
    \includegraphics[width=\textwidth]{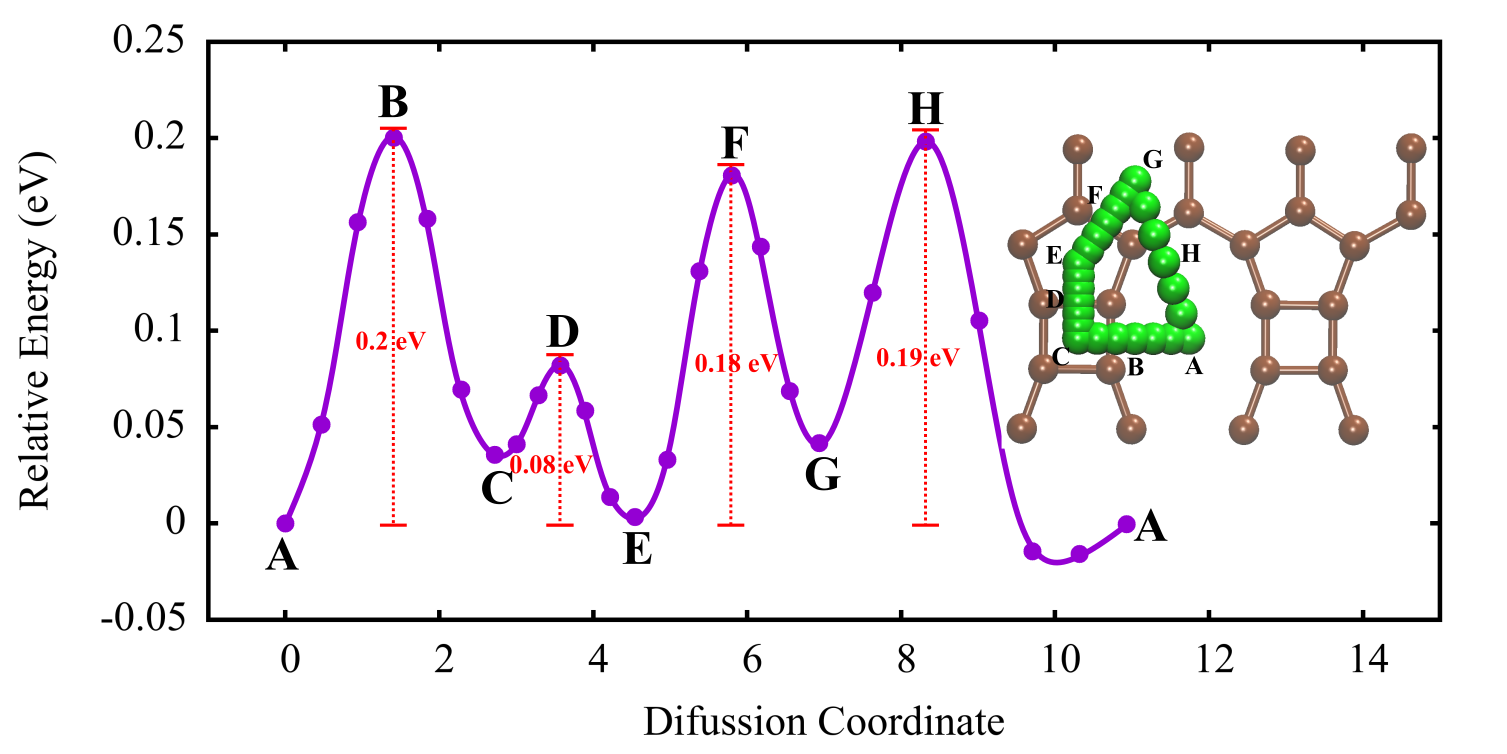}
    \caption{Diffusion energy barriers of the Lithium atom on the surface of TPDH-gr. Inset figure: all intermediate Li images for each diffusion path are shown.}
    \label{fig:difusion_barriers}
\end{figure}

In Figure \ref{fig:2a}, we present the electronic band structure and the density of electronic states for the final configuration of TPDH-gr with a lithium atom adsorbed in the D region. As can be seen, the TPDH + Li system does not have a band gap because of the intrinsic metallic nature of TPDH-gr. To gain a deeper understanding of the adsorption properties between the Li atom and TPDH-gr, we performed calculations of the charge density differences for the D adsorption region, as shown in Figure \ref{fig:2b}. Our analysis revealed that electrons mainly accumulated around the carbon atoms, a result attributed to the higher electronegativity of carbon compared to lithium.\\

The adsorption energy values $E_{ads}^{bcc}$ for Li atoms (N = 1,2,...,6) are presented in Figure \ref{fig:3a}, where we connect the most stable configurations with a dotted line. It is clear that N=3 is the optimal amount of Li atoms adsorbed on TPDH-gr, with a maximum value of $E_{ads}^{bcc} \sim -1.0$ eV. In Figure \ref{fig:3a}, it can also be seen that the maximum amount of Li that can be stored is N=6, which only allows a maximum of 7 unique configurations with Li atoms separated by $> 2.8$ \AA \ (see inset in Figure \ref{fig:3a}). Considering both TPDH-gr sides, we have N=12 as the maximum limit of lithium atoms adsorbed on the TPDH-gr. With these results, it is possible to calculate the theoretical capacity of TPDH-gr (for storage of Li atoms), following equation \ref{eq:capacity}. The C value obtained was 1116 mA h/g. This capacity value is three times larger than the theoretical capacity of graphite (C = 372 mA h / g)\cite{asenbauer2020success}, which is the anode most used in commercial LIB, and is comparable to the theoretical capacity of several 2D carbon allotropes: graphenylene (930 mA h/g), popgraphene (1487 mA h/g) and pentagraphene (1489 mA h/g) \cite{Rajkamal2019}.

An important factor regarding the charge-discharge performance of a LIB is the diffusion of Li atoms on the surface of TPDH-gr. In Figure \ref{fig:difusion_barriers}, the Li diffusion barriers for the four diffusion pathways considered in this study (see Figure \ref{fig:1b}) are shown, which were calculated using the NEB method. As can be observed in Figure \ref{fig:difusion_barriers}, the energy barriers for Li diffusion have small values, with 0.2 eV being the maximum value for diffusion through pathway 1. Furthermore, the diffusion barriers for pathways 2, 3, and 4 with respect to pathway 1 have the following values: 0.08, 0.18, and 0.19 eV, respectively. The diffusion barrier values for some 2D carbon allotropes are as follows: graphene 0.33 eV, graphite 0.47 eV, biphenylene 0.68 eV, and Pop-graphene 0.58 eV \cite{Rajkamal2019}. Therefore, because of the small diffusion barrier value of TPDH-gr, we can deduce that it could potentially exhibit equal or better performance in the charge and discharge process compared to those of graphene and graphite, which are currently the most widely used anodic materials for LIBs.\\

For a better understanding of the Li diffusion process on TPDH-gr, we calculated the diffusion coefficients for the four selected pathways. To achieve this, the Arrhenius equation was employed \cite{Gao2023}:
\begin{equation}
    D_{\text{coeff}}(T) = L^{2} \nu_{0} \text{ exp}\left( - \frac{\Delta E_{b}}{k_{B}T} \right)
    \label{eq:coeff}
\end{equation}
where $\Delta E_{b}$ is the value of the diffusion barriers, $L$ is the length of the diffusion path for the Li atom, $T$ is the absolute temperature in Kelvin units, $\nu_{0}$ is the vibration frequency, which usually has a value of 10 THz, and $k_{B}$ is the Boltzmann constant ($8.62$ $\times 10^{-5}$ eV/K) \cite{Gao2023}.\\

\begin{figure}[h]
    \centering
    \includegraphics[scale=0.9]{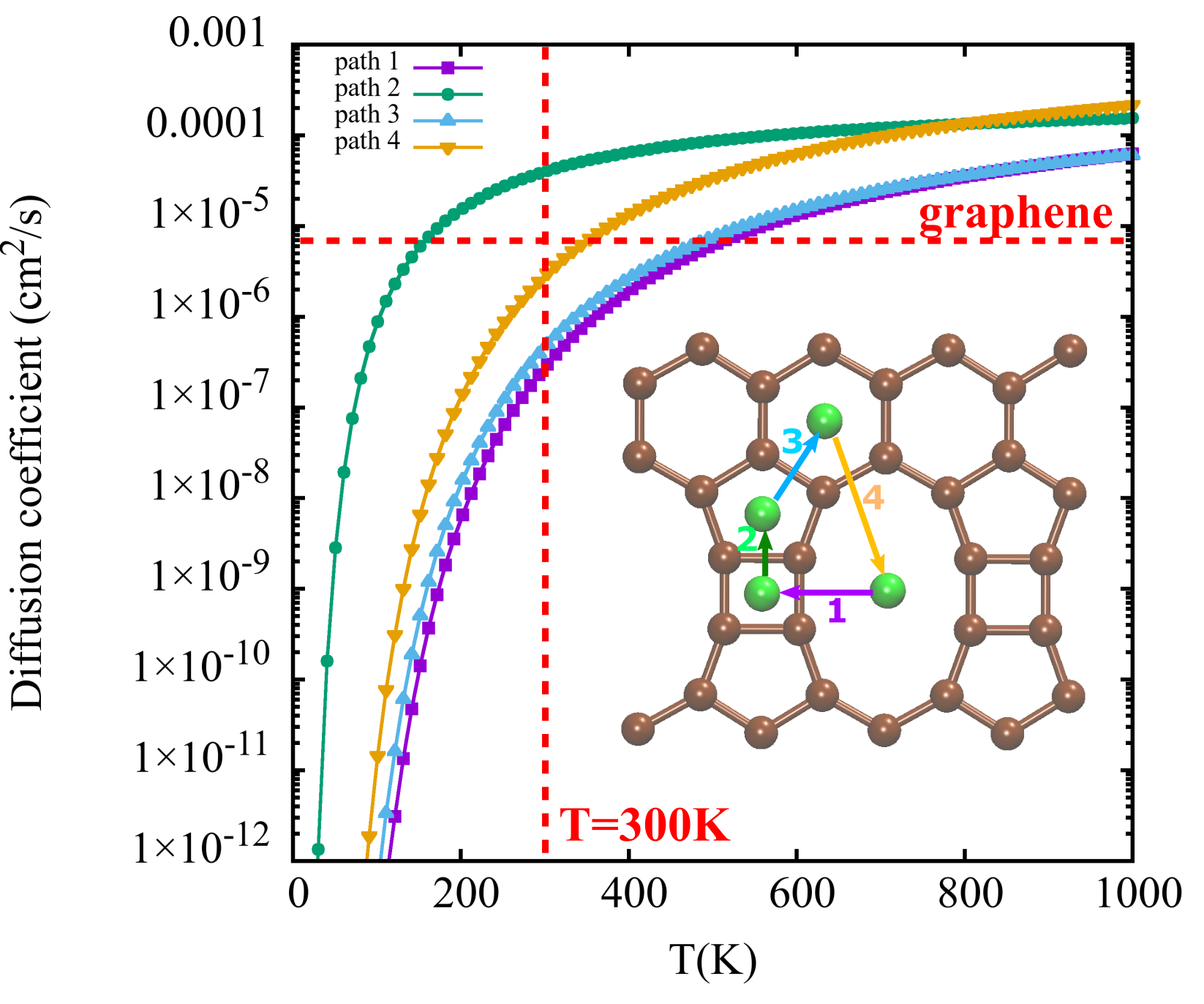}
    \caption{Diffusivity of Li atom over a TPDH-gr as a function of temperature (T). Inset figure: The four diffusion paths of Li on the TPDH-gr are shown.} 
    \label{fig:diffusion_coef}
\end{figure}

In Figure \ref{fig:diffusion_coef}, the diffusion coefficient is shown as a function of global temperature for the Li diffusion pathways in TPDH-gr. As observed, the $D_{\text{coeff}}$ values at T=300K vary, with pathways 1 and 3 showing slower Li diffusion. In contrast, the $D_{\text{coeff}}$ of Li in pathways 2 and 4 have larger values, indicating faster diffusion in TPDH-gr, even faster than diffusion in graphene ($6 \times 10^{-6} cm^{2}/s$) \cite{Gao2023}.

\begin{figure}[h]
\centering
\begin{subfigure}{0.49\textwidth}
    \includegraphics[width=\textwidth]{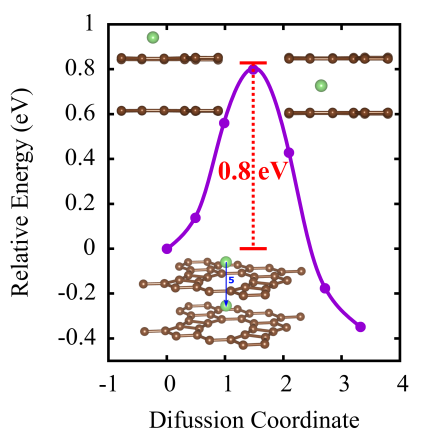}
    \caption{}
    \label{fig:barrier_bilayer}
\end{subfigure}
\hfill
\begin{subfigure}{0.49\textwidth}
    \includegraphics[width=\textwidth]{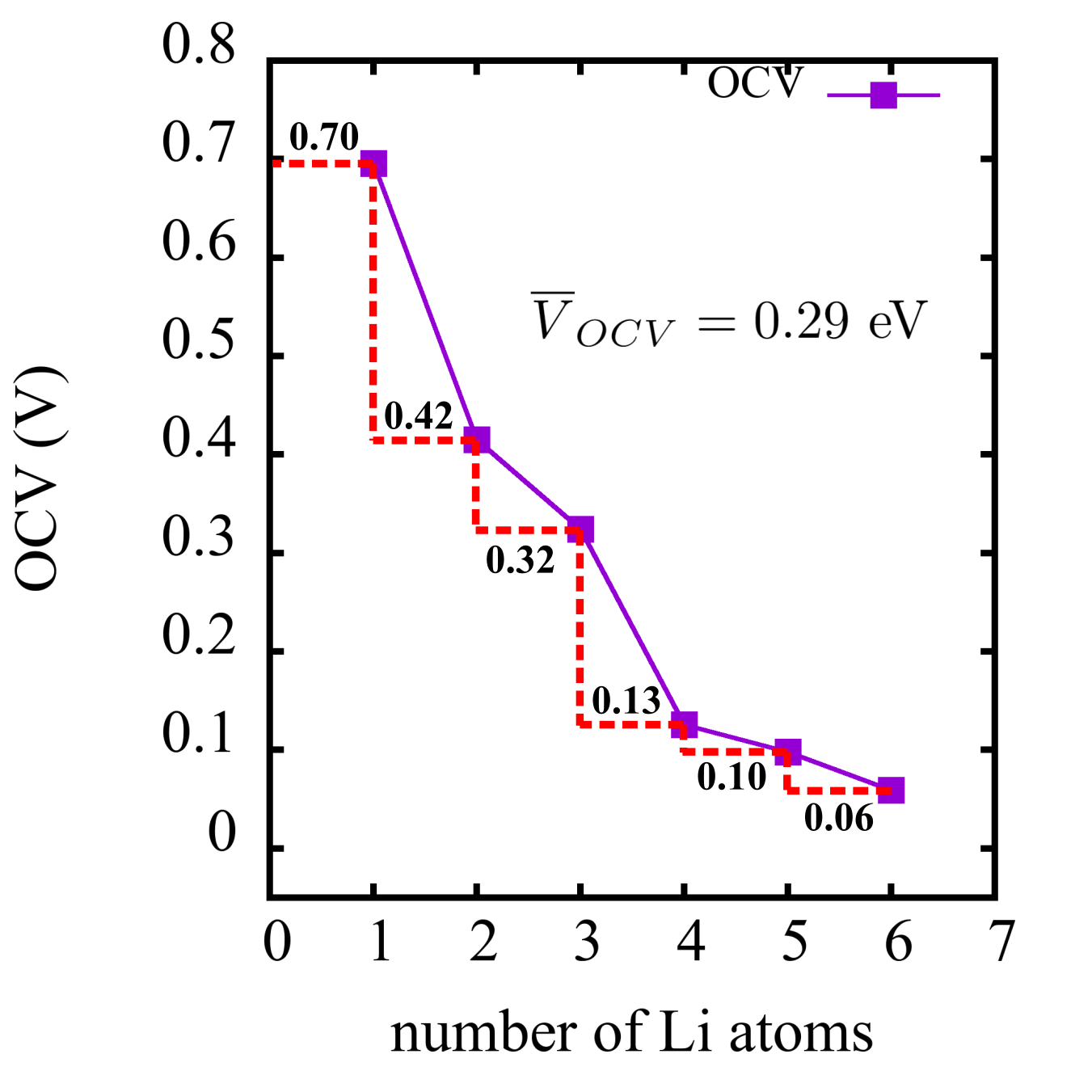}
    \caption{}
    \label{fig:ocv_plot}
\end{subfigure}
\caption{(a) Diffusion barrier of the Lithium atom into a TPDH-gr bilayer. These interlayer diffusions were considered only in the (D) region. (b) Open circuit voltage (OCV) as a function of the number (L) of Li atoms adsorbed on TPDH-gr.}
\label{fig:figures}
\end{figure}

The Li diffusion barrier for a bilayer TPDH-gr is shown in figure \ref{fig:barrier_bilayer}, where the diffusion path corresponds to the migration of the Li atom from the top (region D) to the middle of TPDH-gr, as can be seen in Figure \ref{fig:1c}. The diffusion barrier has a value of 0.8 eV, which indicates that Li diffusion occurs more easily on the surface of the TPDH-gr than along the diffusion path between layers. In Figure \ref{fig:ocv_plot}, the OCV is presented as a function of the number of Li atoms adsorbed in TPDH-gr. As observed in Eq. \ref{eq:ocv}, the OCV correlates with the value of the adsorption energy of the Li atoms. Thus, small OCV values indicate moderate Li adsorption, allowing for easy Li desorption. The average OCV value shown in Figure \ref{fig:ocv_plot} is 0.29 V, which is very close to the OCV values of other 2D carbon allotropes considered for application as anodes in LIBs: Pop-graphene (0.30 V) and graphenylene (0.34 V) \cite{Ullah2024}.

\section{Conclusions}\label{sec4}
In conclusion, we have explored the potential of TPDH-gr as an anode in LIB using DFT-based calculations. Our results indicate that Li-atom adsorption on TPDH-gr is moderate (around 0.70 eV), allowing easy desorption. Additionally, TPDH-gr has a high theoretical capacity of 1116 mAh/gr, making it one of the few 2D carbon allotropes with metallic properties and significant theoretical capacity. Moreover, energy barrier, diffusion coefficient, and OCV calculations show rapid Li-atom diffusion on the surface, stable intercalation of Li atoms, and good performance during the charge and discharge cycles of the LIB. Together, these findings present TPDH-gr as a viable candidate for anode material in LIB.

\section{Acknowledgements}

JGQ thanks the UFABC Multiuser Computational Center (CCM-UFABC) for the computational resources provided. DSG thanks the Center for Computational Engineering \& Sciences (CCES) at Unicamp for financial support through the FAPESP/CEPID Grant 2013/08293-7 and PASA to CNPq (Grant 308428/2022-6). PASA and BI thank to CNPq - INCT (National Institute of Science and Technology on Materials Informatics, grant n. 371610/2023-0).

\bibliographystyle{unsrt} 
\bibliography{references}
\end{document}